\documentstyle[preprint,aps,epsf,floats]{revtex}
\def\vub{V_{ub}}
\def\as{\alpha_s}
\def\ba{\begin{eqnarray}}
\def\ea{\end{eqnarray}}
\begin{document}

\preprint{
\vbox{\halign{&##\hfil    \cr
        & CMU-02-01 \cr
        &  FERMILAB-Pub-01/062-T \cr
        }}}

\title{A Comment on the Extractions of $V_{ub}$  from Radiative Decays}

\author{Adam K.\ Leibovich$^1$, Ian Low$^2$, and I.\ Z.\ Rothstein$^2$}

\address{
$^1$ Theory Group, Fermilab, P.O. Box 500, Batavia, IL 60510 \\
$^2$ Department of Physics,
         Carnegie Mellon University,
         Pittsburgh, PA 15213}

\maketitle

{\tighten
\begin{abstract}
We present a model independent 
closed form expression
for $|V_{ub}|^2/|V_{tb} V_{ts}^*|^2$, which includes
the resummation of large endpoint logarithms as well as the
interference
effects from the operators $O_2$ and $O_8$. 
We demonstrate that the method to extract $|V_{ub}|$ presented by the authors
in hep-ph/9909404, and modified in this letter to include interference
effects, is not just a refinement of the method introduced in 
hep-ph/9312311. We also discuss the model dependence of the latter proposal.
Furthermore, we show that 
the resummation is not negligible and that the Landau pole
does not  introduce any significant uncertainties. 

\end{abstract}
}


\newpage

Testing the Standard Model in the
Cabbibo-Kobayashi-Maskawa sector has been hindered by the
relatively large uncertainties in the matrix element $V_{ub}$. The
absolute value of this matrix element has been extracted by the study
of inclusive charmless $B$ decays, with  large uncertainties from
model dependence. There really is no way to define
a theoretical error in this extraction, 
since the calculations are not based on
a controlled expansion. The model dependence is introduced as a
consequence of the need to make a cut on the electron energy spectrum
near the endpoint to eliminate the large
background from charmed decays.
This probing of the endpoint region makes the cut rate sensitive
to the Fermi motion of the heavy quark inside the hadron. 
In the past, one has needed to use models for the Fermi motion
leading to the aforementioned uncontrolled errors. It is now well
known that it is possible to avoid the model dependence by
using the data from radiative decays to eliminate the dependence
of the Fermi motion.

In this note we will discuss two proposals for implementing this
idea. One, introduced by Neubert \cite{N1}, and the other
by the authors \cite{us} based on ideas of
Korchemsky and Sterman \cite{KS}.
We will show that the results in 
 \cite{us} are not just
a refinement of Neubert's proposal, which 
is model dependent, whereas the results of \cite{us}
are not. We will further demonstrate that 
 there is a well defined prescription
to handle the Landau singularity which is unambiguous. Finally, we will
 show that, when using the present experimental cut, 
the effect of resummation is not negligible.

Let us first review Neubert's  proposal \cite{N1}, as recently updated
to include interference effects in
\cite{newN}. At {\it tree} level the decay rate near the endpoint may
be written as \cite{N1,Bigi}
\begin{equation}
\frac{d \Gamma}{dx}=\frac{ G_F^2 |V_{ub}|^2 m_b^5}{96 \pi^3}
\left[ F(x)\theta(1-x)+F(1)S(x) \right],
\end{equation}
where $x=2 E_e/m_b$, and $F(x)\approx F(1)$ near the endpoint.  This
result follows from taking the imaginary part of the tree level
current-current correlator.  At leading order in $\Lambda/m_b$, we may
write
\begin{equation}
\theta(1-x) +S(x)=\langle B | \theta(1-x+ i n \cdot \hat{D}) | B\rangle,
\end{equation}
where $n_\mu$ is a light-like  vector satisfying $n\cdot v=1$, and
$\hat{D}^\mu=D^\mu/m_b$.
The photon spectrum in radiative decay may 
similarly be written, also at tree
level,
as 
\begin{equation}
\frac{d \Gamma^\gamma}{dx}=\frac{G_F^2\, \alpha\, m_b^5\, C_7^2}
{32 \pi^4} |V_{tb} V^*_{ts} |^2 
\langle B | \delta(1-x+ i n\cdot \hat{D}) | B\rangle.
\end{equation}
Then using the relation
\begin{equation}
\int^\infty_x dx^\prime (x^\prime -x)\langle B | \delta(1-x^\prime+i n\cdot
\hat{D})| B \rangle
 =\int^\infty_x dx^\prime \langle B |
\theta(1-x^\prime+i n\cdot \hat{D})| B \rangle,
\end{equation}
one can write
\begin{equation}
\label{res}
\left|\frac{V_{ub}}{V_{tb} V^*_{ts}}\right| ^2= \frac{3 \alpha}{\pi} 
|C_7|^2 \frac{\Gamma_u(E_c)}{\Gamma_s(E_c)}+{\cal O}(\alpha_s)+
 {\cal O}(\Lambda/m_b),
\end{equation}
where $\Gamma_i(E_c)$ is the cut  integrated rate.
To take into account the perturbative corrections, the author of
\cite{N1} adds a correction factor $\eta_{\rm QCD}$. 
In \cite{N1} $\eta_{\rm QCD}$ is given by
\begin{equation}
\eta_{\rm QCD}=1+\frac{2 \alpha_s}{9 \pi}\left( 5 \log(r)+\pi^2-
\frac{35}{4}\right).
\end{equation} 
The quantity $r$ is unknown, and depends upon the non-perturbative
structure function. This structure function dependence arises because
it is not truly possible to cancel off the soft effects in this way, once the
radiative corrections are included, because these two effects
are convoluted.

\begin{figure}[t]
\centerline{\epsfysize=10truecm  \epsfbox{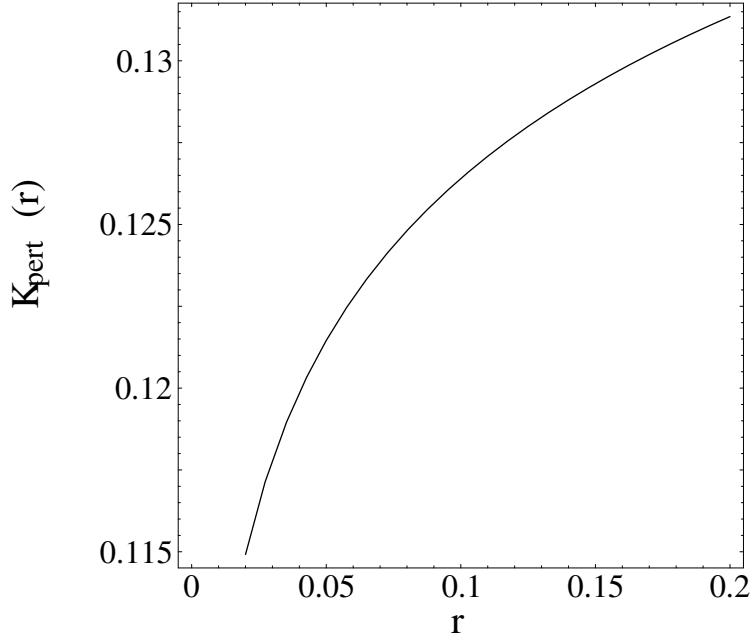} }
\caption[]{\it $K_{\rm pert}$ as a function of the non-perturbative
parameter $r$ in the range $0.02 < r < 0.2$.}
\label{rplot}
\end{figure}

However,  Neubert derived the following bound
\begin{equation}
-\log(r)>-\log\left(1-x_B^c\right).
\end{equation}
While this bound is helpful, it does not really tell us 
much about the relative size of the model dependence. Varying
$r$ within its allowed range can significantly change the 
radiative corrections. In Fig.~\ref{rplot} we plot the parameter
$K_{\rm pert}$ defined in Eq.~(3) of Ref.\cite{newN}, which updates
$\eta_{\rm QCD}$ by including interference effects,
 as a function of $r$. It is clear that $K_{\rm pert}$ is quite sensitive
to the value of $r$ and, unfortunately, a priori we have no idea what value
of $r$ to choose.

The proposal of Ref.~\cite{us}, on the other hand, has no model
dependence. The calculations, based on the factorization shown by
Korchemsky and Sterman \cite{KS} and the results of \cite{AR}, 
lead to
\begin{eqnarray}
\label{oneforexp}
\frac{|V_{ub}|^2}{|V_{ts}^* V_{tb}|^2} &=&
\frac{3\,\alpha\,|C_7(m_b)|^2}{\pi}
   \int^1_{x_B^c}dx_B\frac{d\Gamma}{dx_B}\nonumber\\
&&\times\,\left\{\int^1_{x_B^c}dx_B \int_{x_B}^{1} du_B\;u_B^2
   \,\frac{d\Gamma^\gamma}{du_B}\,
   K\left[x_B;\frac4{3\pi\beta_0}\log(1-\alpha_s\beta_0\,l_{x_B/u_B})\right]
   \right\}^{-1}\;, 
\end{eqnarray}
where the expression for $K$ can be found in \cite{us} and
$l_{x/u} = -\log[-\log(x/u)]$. $x_B^c$ is the larger of the two
energy cuts for the electron energy spectrum of $B\to X_u e \nu$ and 
the photon energy spectrum of $B \to X_s\gamma$. 
In addition to including the full ${\cal O}(\alpha_s)$ corrections, 
this result also includes a summation of the next-to-leading Sudakov 
logarithms $(\log(1-x_B^c))$
which become large as $x_B^c$ approaches one.
This result may be re-written as 
\begin{eqnarray}
\label{forexp}
\frac{|V_{ub}|^2}{|V_{ts}^* V_{tb}|^2} &=&
\frac{3\,\alpha\,C_7(m_b)^2}{\pi}
 \int^1_{x_B^c}dx_B\frac{d\Gamma}{dx_B}\times\,\left\{\int^1_{x_B^c}du_B W[u_B,x_B^c]
   \,\frac{d\Gamma^\gamma}{du_B}\,
   \right\}^{-1} ,\\
\label{weight}
W[u_B,x_B^c] &=& u_B^2 \int_{x_B^c}^{u_B} dx_B \;
K\left[x_B;\frac4{3\pi\beta_0}\log(1-\alpha_s\beta_0\,l_{x_B/u_B})\right],
\end{eqnarray}
where $W[u_B,x_B^c]$ is a weighting function which is approximately linear.

Next we would like to address the issue of the Landau pole.  The
argument of $K$ diverges when $1-\alpha_s\beta_0\,l_{x_B/u_B}=0$. In
the denominator of Eq.~(\ref{forexp}) the integration region is a
triangular region bounded by $x_B^c\le x_B\le u_B\le 1$.  
The Landau pole is located at $(x_B/u_B)_{max} = 1 -
\exp[-1/(\alpha_s\beta_0)] \approx 0.999$. 
One way to avoid the pole is to integrate over the
region $x_B \le \rho\, u_B$, where $\rho \lesssim 0.999$. Since the
physical radiative rate is a smooth function, the area we cut off from
the integration region should not incur substantial error in the
extraction of $|\vub|$.  When cutting the integration region as
described here, the weight function remains approximately linear.

\begin{figure}[t]
\centerline{\epsfysize=13truecm  \epsfbox{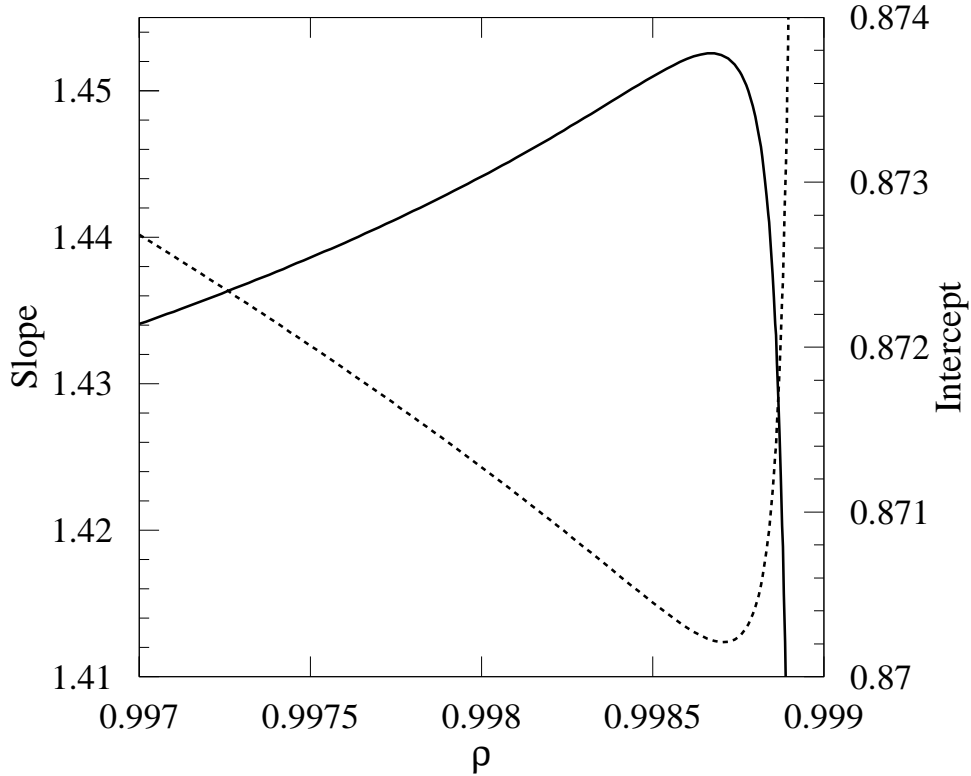} }
\caption[]{\it The slope (solid line) and $x$-axis intercept (dotted
line) of the weight function as a function of $\rho$ for $x_B^c=0.87$
and $\as=0.21$.}
\label{slopeintercept}
\end{figure}

However, an important question in practice is how close we can get to
the Landau pole region \cite{Low}. This question arises because, as we
get very close to the Landau pole, the perturbative resummation breaks
down.  Ideally, we would like to cut off as little integration region
as possible while still leaving a well-behaved perturbative
resummation. After the introduction of $\rho$, Eqs.~(\ref{forexp}) and
(\ref{weight}) become
\begin{eqnarray}
\label{vubweight}
\frac{|V_{ub}|^2}{|V_{ts}^* V_{tb}|^2} &=&
\frac{3\,\alpha\,|C_7(m_b)|^2}{\pi}
\int^1_{x_B^c}dx_B\frac{d\Gamma}{dx_B}\,
\times\,\left\{\int^1_{x_B^c/\rho}du_B
\,\frac{d\Gamma^\gamma}{du_B}\,W[u_B,x_B^c,\rho] \right\}^{-1} ,\\
\label{w2}
W[u_B,x_B^c,\rho] &=& u_B^2 \int_{x_B^c}^{\rho\,u_B} dx_B \;
K\left[x_B;\frac4{3\pi\beta_0}\log(1-\alpha_s\beta_0\,l_{x_B/u_B})\right].
\end{eqnarray}
To determine the optimal value of $\rho$, we plot in
Fig.~\ref{slopeintercept} the slope and $x$-axis intercept of the
weight function, Eq.~(\ref{w2}), for various values of
$\rho$. It is clear that, as $\rho$ varies from 0.97 to 0.998, $W$
converges to an asymptote.  However, for $\rho\sim0.9988$, the
perturbative expansion breaks down, as is evident from the abrupt
change in the behavior of the curves.  The weight function changes
abruptly, which signals the breakdown of perturbative resummation.
 
As $\rho$ gets smaller, we are cutting off more of the integration region,
which results in a weight function with different intercepts and slopes, 
as shown
in Fig.~\ref{slopeintercept}. However, we have to bear in mind that
this is an approximation scheme. The more we cut off the integration
region, the worse an approximation it is. Therefore, it is not
at all surprising that different values of $\rho$ yields different
values of $|V_{ub}|$ when using Eq.~(\ref{vubweight}). Ideally we would
like $\rho$ to be as close to unity as possible, in order to have a
good approximation, while maintaining a controlled resummation. It is
clear from Fig.~\ref{slopeintercept} that the true weight function is
approached asymptotically. When performing the analysis
experimentally, we can either use $\rho=0.9987$, or try to extrapolate
$W$ all the way up to $\rho=1$. The difference should be well within
the theoretical error.

\begin{figure}[t]
\centerline{\epsfysize=13truecm  \epsfbox{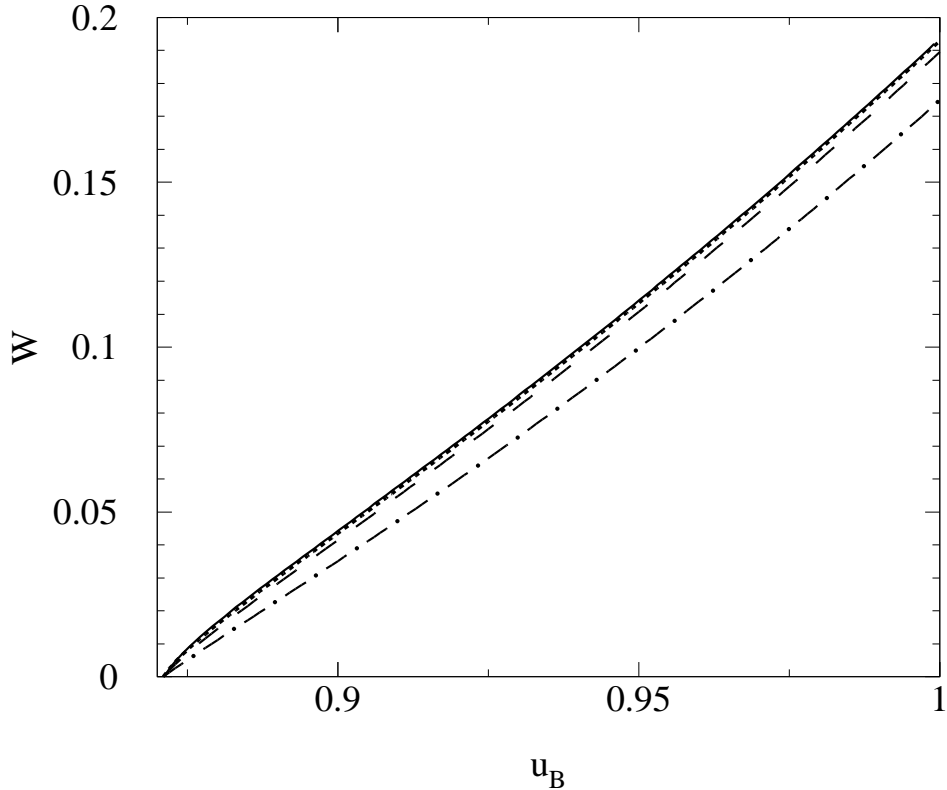} }
\caption[]{\it Weight function obtained by expanding the argument of
$K$ in Eq.~(\ref{weight}) to different powers of $\alpha_s$, using 
$x_B^c=0.87$ and $\as=0.21$. The
dot-dashed line is expanding to order $\alpha_s$, the dashed line to
order $\alpha_s^3$ and the dotted line to $\alpha_s^5$.  The weight
function is quickly converging to the solid line, which is the weight
function from Eq.~(\ref{vubweight}) using $\rho=0.9987$.}
\label{prescription}
\end{figure}

Another prescription for avoiding the Landau pole is to expand the
second argument of $K$ in Eq.~(\ref{weight}) as a power series in
$\alpha_s$,
\begin{equation}
K\left[x,\frac{4}{3 \pi\beta_0} 
  \log(1-\alpha_s\beta_0 l_{x_B/u_B})\right] = 
K\left[x,-\frac{4}{3 \pi} 
  \left(\alpha_s l_{x_B/u_B} 
  + \frac12 \alpha_s^2 \beta_0 l_{x_B/u_B}^2 + \cdots \right)\right] .
\end{equation}
This corresponds to expanding $g_{sl}$ of Ref.\cite{us} in the 
exponent.\footnote{Note that expanding in the exponent is not
equivalent to expanding in $\alpha_s$. Indeed an expansion in
$\alpha_s$ (i.e., expanding $K$ as a series in $\alpha_s$) leads to a
very poorly behaved series. The beauty of the resummation is that the
series is reorganized in such a way that the expansion in the exponent
is well behaved \cite{LR}.}  We can check the convergence of this
prescription by expanding to different orders in $\alpha_s$.  In
Fig.~\ref{prescription}, we expand the argument of $K$ to orders
$\alpha_s$ (dot-dashed line), $\alpha_s^3$ (dashed line) and
$\alpha_s^5$ (dotted line).  We also show (solid line) the weight
function from Eq.~(\ref{w2}) using $\rho=0.9987$.  It is clear that
the expansion is quickly converging, and it is converging to the
weight function using the other prescription.  It is therefore evident
that there is an unambiguous choice of weighting function which can be
used, with negligible error introduced.

\begin{figure}[t]
\centerline{\epsfysize=14truecm  \epsfbox{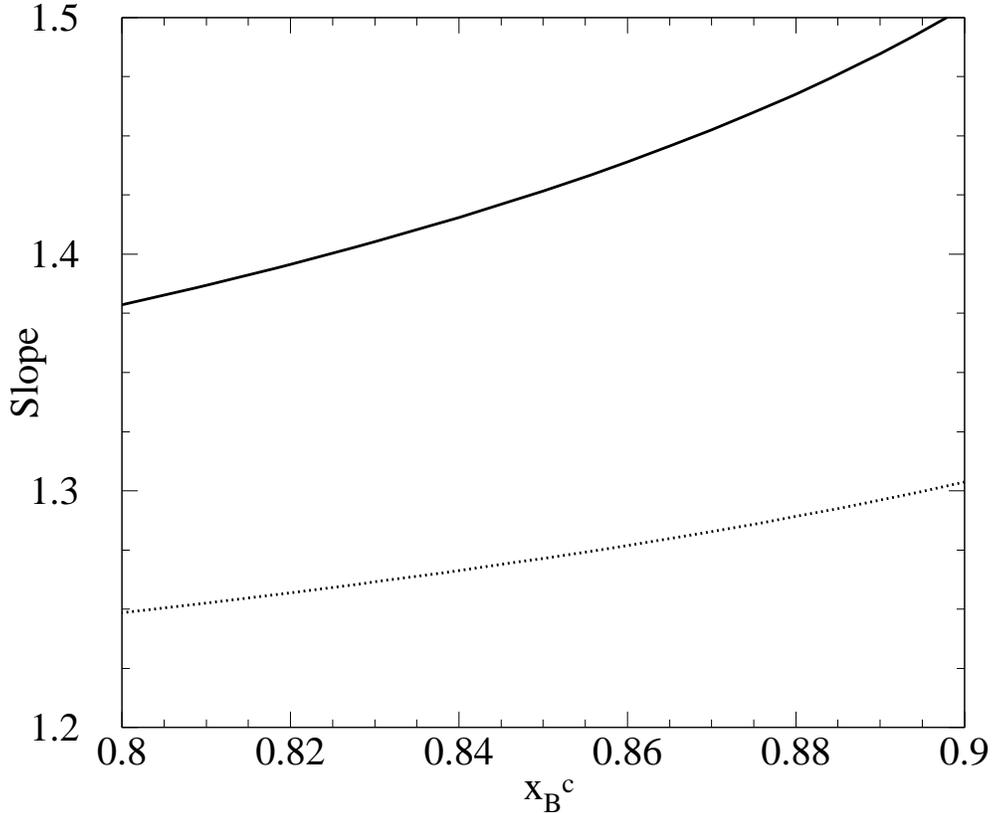} }
\caption[]{\it The slope of the weight function 
as a function of the cut showing the effects of resummation. 
The dotted line is the slope without
resumming the Sudakov logarithms.}
\label{newslope}
\end{figure}

Now we would like to discuss the effect of resummation.\footnote{For
other work on resumming endpoint logs see \cite{Ugo,bauer}.} 
Since the weight
function in Eq.~(\ref{weight}) is approximately linear, we plot
in Fig.~\ref{newslope} the slope of the weight function with the fully
resummed result versus the slope without resumming the Sudakov logarithms,
with the choice of $\rho=0.9987$. We see that the resummation has 
roughly a 10\% effect on the slope of the weight function, for the current
experimental cut on the electron energy spectrum, $E_{cut}=2.3\ {\rm GeV}$
or $x_B^c=0.87$. In our original paper, Ref.\cite{us}, we proposed to use
$\rho=0.99$. However, at that time we did not fully investigate 
 the sensitivity due to the
Landau pole. Had we used the choice $\rho=0.99$ in Fig.~\ref{newslope}, we
would have found that the resummation has a very small effect. This is because
we would have cut off a region where the Sudakov logarithms are important. 
Now it should
be clear that, when the optimal value of $\rho$ is used, the resummation does
have a non-negligible effect.

Finally, it was correctly pointed 
out in \cite{newN} that we mistakenly neglected the
contribution from interference terms which can be large when studying
the integrated radiative decay rate.
At leading order, 
the only operator that is important is $O_7$, the electromagnetic
penguin operator.  At order $\alpha_s$ in the decay rate, $O_7$
interferes
 with $O_2$ and $O_8$ \cite{GHW}.
The contribution from other operators are small and can be neglected.
The contribution from $O_2O_7$ and $O_7O_8$ terms are also 
suppressed by exponentiated Sudakov
logarithms, and can be included trivially in our
formula by changing the overall factor in Eq.~(\ref{forexp}) or 
Eq.~(\ref{vubweight}) to\footnote{When using the 
hadronic mass spectrum to extract $V_{ub}$\cite{had}, we should take
into account the interference effect in a similar fashion.} 
\begin{equation}
\frac{|V_{ub}|^2}{|V_{ts}^* V_{tb}|^2} =
\frac{3\,\alpha\,C_7^{(0)}(m_b)^2}{\pi}(1+H^\gamma_{mix})
 \int^1_{x_B^c}dx_B\frac{d\Gamma}{dx_B}\times\,\left\{\int^1_{x_B^c}du_B W(u_B)
   \,\frac{d\Gamma^\gamma}{du_B}\,
   \right\}^{-1} ,
\end{equation}
where 
\begin{equation}
\label{newfactor}
H^\gamma_{mix} = \frac{\alpha_s(m_b)}{2\pi C_7^{(0)}}\left[
C_7^{(1)} + C_2^{(0)} \Re(r_2) + 
C_8^{(0)} \left(\frac{44}9 - \frac{8\pi^2}{27}\right)\right].
\end{equation}
In Eq.~(\ref{newfactor}), all the Wilson coefficients,  evaluated at
$m_b$, are ``effective'' as
defined in \cite{CMM},  and 
$\Re(r_2) \approx -4.092 + 12.78(m_c/m_b - 0.29)$ \cite{GHW}.
The numerical values of the Wilson coefficients are\cite{Kagan:1999ym}: 
$C_2^{(0)}(m_b) \approx 1.11,\ C_7^{(0)}(m_b)\approx -0.31,\ 
C_7^{(1)}(m_b)\approx 0.48$, and $C_8^{(0)}(m_b)\approx -0.15$. 
With this expression in hand we believe 
it to be relatively straightforward to
extract $|V_{ub}|^2$  with theoretical errors on the
order of $\Lambda/m_b$.

\acknowledgments 
We would like to thank Zoltan Ligeti and Matthias Neubert for useful 
discussions. This work was supported in part by the Department of Energy 
under grant number DOE-ER-40682-143 and DE-AC02-76CH03000. A.K.L. would like
to thank the theory group at CMU for its hospitality.

\tighten


\end{document}